\documentclass[amsmath,prl, twocolumn, nofootinbib,superscriptaddress]{revtex4}
\usepackage{graphicx}
\usepackage{bbold}
\usepackage{color}
\newcommand{\be}{\begin{eqnarray}}
\newcommand{\ee}{\end{eqnarray}}

\newcommand{\nn }{\nonumber}

\begin{document}
\title{ Variational Scheme to Compute Protein Reaction Pathways using Atomistic Force Fields with  Explicit  Solvent}
\author{ S. a Beccara}
\affiliation{European Centre for Theoretical Nuclear Physics and Related Areas (ECT*-FBK), Strada delle Tabarelle 287, Villazzano (Trento), 38123 Italy}
\affiliation{Trento Institute for Fundamental Physics and Applications (INFN-TIFPA), Via Sommarive 14 Povo (Trento), 38123 Italy. }
\author{L. Fant}
\affiliation{Dipartimento di Fisica, Universit\`a degli Studi di Trento, Via Sommarive 14 Povo (Trento), 38123 Italy. }

\author{P. Faccioli\footnote{faccioli@science.unitn.it}}
\affiliation{Dipartimento di Fisica, Universit\`a degli Studi di Trento, Via Sommarive 14 Povo (Trento), 38123 Italy. }
\affiliation{Trento Institute for Fundamental Physics and Applications (INFN-TIFPA), Via Sommarive 14 Povo (Trento), 38123 Italy. }
\begin{abstract}

We introduce a variational approximation to the microscopic dynamics of rare conformational transitions of macromolecules.  Within this framework it is possible to simulate on a small computer cluster  reactions as complex as protein folding, using  state of the art all-atom force fields in explicit solvent.  We test this method against molecular dynamics (MD) simulations of the folding of an $\alpha$- and  a $\beta$-protein  performed with the same all-atom force field on the \emph{Anton} supercomputer. We find that our approach yields results  consistent with those of MD simulations,  at a computational cost orders of magnitude smaller. 

\end{abstract}
\maketitle

The development of the special-purpose \emph{Anton} supercomputer  has recently opened the way to MD simulations of  bio-molecules consisting of several hundreds atoms, covering time intervals in the millisecond range~\cite{DESRes1}. By using this facility, Shaw and co-workers characterized the reversible folding of several small proteins, showing that the existing all-atom force fields are able to attain the correct protein native structures \cite{DESRes1, DESRes2, DESRes3}. 
Unfortunately, many biologically important conformational reactions occur at time scales many orders of magnitude larger than the millisecond. Hence, it is important to continue the development of more efficient algorithms to sample the reactive pathways space (see e.g. \cite{PSreview} and references therein).

In particular, in the  Dominant Reaction Pathways (DRP) approach \cite{Elber, DRP, Doniach, aBeccara2012},
 microscopic trajectories $X(\tau)$, connecting given initial and final molecular configurations $X_i=X(0)$ and $X_f=X(t)$ are determined by maximizing their probability density
$\mathcal{P}[X]$ in the Langevin dynamics. This algorithm was first  validated against MD using both simplified and realistic atomistic force fields (see e.g. Ref. \cite{aBeccara2012}).  Next, it was applied  to characterize in atomistic detail conformational reactions  far too slow to be investigated by means of plain MD. Notable examples include the folding of a knotted protein~\cite{aBeccara2013} and the latency transition of several serpins \cite{serpin}.

One crucial limitation of the DRP method is that it can only be applied in \emph{implicit solvent} simulations. In this work we overcome this limitation by introducing a new variational approximation suitable also for atomistic simulations in \emph{explicit solvent}. 

\begin{figure}[t!]
\includegraphics[width= 9cm]{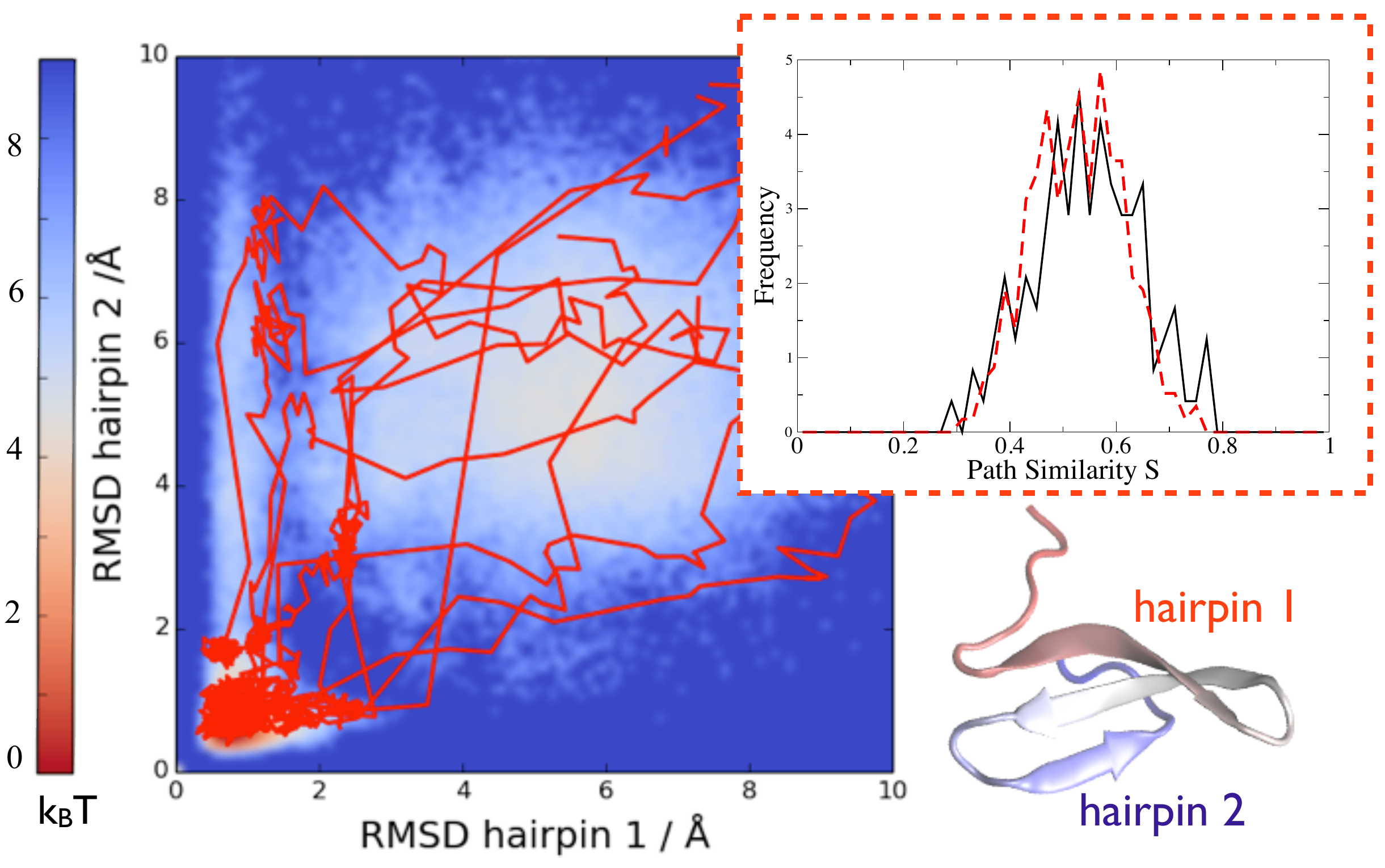}
\caption{Folding trajectories for the WW-domain (crystal structure shown in the bottom right corner) obtained in the variational approximation projected on the plain defined by RMSD to the native structure of the two hairpins. The color-map in  the background represents the free-energy landscape obtained from the frequency histogram of the Anton MD trajectories. In the top-right corder: The similarity distribution between variational and MD folding pathways (dashed line) compared with the intrinsic similarity of MD folding pathways (solid line).  }
\label{Fig1}
\end{figure}
Let $(X,Y)$ represent a point the system's configuration space, where  $X=({\bf x}_1, ..., {\bf x}_N)$ and $Y=({\bf y}_1, ..., {\bf y}_{N'})$ denote the solute and solvent coordinates, respectively. The Langevin equations  for the solvent and solute are
\be
\label{Lang}
m_i \ddot {\bf x}_i &=& -m_i \gamma_i \dot{\bf x}_i - \nabla_i U +  \eta_i(t)\nn\\
m_j \ddot {\bf y}_j &=& -m_j \gamma_j \dot{\bf y}_j - \nabla_j  U + \eta_j(t),
\ee
where $U(X,Y)$ is the potential energy, $\eta_i$ is a white noise and $m_i$ and $\gamma_i$ denote mass and viscosity, respectively. 

We are interested in the probability density for the solute to make a transition from $X_i$ to $X_f$ in a time $t$, along a given  path $X(\tau)$. This  is given by the path integral (PI) 
\be\label{unbiased}
\mathcal{P}[X]
&=&\int\mathcal{D}Ye^{-S_{OM}[X,Y]-\frac{U(X_i, Y_i)}{k_B T}},
\ee where $S_{OM}[  X,Y]$ is the  Onsager-Machlup functional, to be defined below. 
Maximizing $\mathcal{P}[ X]$ with respect to the path $ X$ yields the DRP optimum condition~\cite{Elber, DRP, Doniach}: 
%\be\label{var}
$
\frac{\delta}{\delta  X} \langle S_{OM}[ X, Y] \rangle_Y =0,$
%\ee 
where the average $\langle \cdot \rangle_Y$ refers to the PI over $Y(\tau)$.  

Unfortunately, computing this average with the accuracy required for the path optimization is computationally unfeasible, because of large statistical fluctuations.  To overcome this problem, we need to derive an optimum criterion which does not involve any average over the solvent dynamics. 

We begin by considering a modified stochastic dynamics, defined by introducing into Eq. (\ref{Lang}) an \emph{external} (possibly time-dependent) biasing force ${\bf F}^{bias}_i(X,t)$, acting  on the solute atoms only and accelerating the transition to the product. 
The probability of a given reactive pathway $ X(\tau)$ in the biased dynamics is given by
\be\label{PI2}
\mathcal{P}_{bias}[ X] =  \int \mathcal{D}Y e^{-S_{bias}[ X,Y]-\frac{U(X_i,Y_i)}{k_BT}},
\ee where 
the functional $S_{bias}[X,Y]$ is defined as \be\label{StrVMD}
&&\hspace{-0.6cm}S_{bias}\equiv \frac{1}{4k_BT}   \int_0^t d\tau~\left[\sum_{i=1}^N \frac{1}{\gamma_i m_i}\left(m_i \ddot{\bf x}_i + m_i \gamma_i \dot{\bf x}_i \right. \right.\hspace{0.3 cm}\\
&&\left.\left.\hspace{-0.5cm} + \nabla_i U  - {\bf F}_i^{bias}\right)^2 + \sum_{j=1}^{N'} \frac{1}{\gamma_j m_j}\left(m_j \ddot{\bf y}_i + m_j \gamma_j \dot{\bf y}_j+ \nabla_j U\right)^2\right]\nn
\ee 
The  Onsager-Machlup functional $S_{OM}[ X, Y]$ entering Eq. (\ref{unbiased}) is recovered setting ${\bf F}^{bias}_i=0$ in Eq. (\ref{StrVMD}).

Let us now return to the problem of computing the reaction pathways in the \emph{unbiased} Langevin dynamics (\ref{Lang}). Using the standard re-weighting trick we can write  the variational condition $\frac{\delta}{\delta  X}\mathcal{P}[ X]=0$ as
\be\label{funder}
\frac{\delta}{\delta  X}\left[ \mathcal{P}_{bias}[ X]~\langle e^{-( S_{OM}[ X, Y]-S_{bias}[ X, Y;t]) }\rangle_{bias}\right]=0.
\ee

We now introduce our main approximation, by restricting the search for the optimum path $ X(\tau)$ within an ensemble of trajectories generated by integrating the \emph{biased} Langevin equation. By definition, these  paths have a large statistical weight in the biased dynamics, i.e. they lie in the functional vicinity of some path $\bar X(\tau)$ which satisfies $\frac{\delta }{\delta \bar X}\mathcal{P}[\bar X]=0$.  Thus, the typical biased paths approximatively satisfy the stationary condition
\be
\label{quasistationary}
\frac{\delta}{\delta X} \mathcal{P}[ X] \simeq 0
\ee  
and obey the corresponding saddle-point equations of motion:
\be
\label{SPE}\begin{cases}
m_i \ddot {\bf x}_i+m_i \gamma_i \dot{\bf x}_i + \nabla_i U-{\bf F}_i^{bias}\simeq 0 & \mbox{}\\
m_j \ddot {\bf y}_j+m_i \gamma_j \dot{\bf y}_j + \nabla_j U\simeq 0.& \mbox{}\\
\end{cases}
\ee
We emphasize that Eq. (\ref{quasistationary}) and (\ref{SPE}) are only satisfied by paths generated by integrating the biased Langevin equation. Using Eq. (\ref{quasistationary}) in Eq. (\ref{funder}) we find
\be\label{funder2}
0\simeq \frac{\delta}{\delta  X} \langle e^{-( S_{OM}[ X, Y]-S_{bias}[ X, Y]) }\rangle_{bias}.
\ee
The crucial point to observe is that, since the biasing force ${\bf F}_i^{bias}$ acts on the solute atoms only, the difference $\Delta S [ X] \equiv S_{OM}[ X,Y]-S_{bias}[ X, Y]$  does not depend on the solvent paths $Y(t)$. Thus, Eq. (\ref{funder2}) reduces to
$
\frac{\delta}{\delta  X} \Delta S[ X]=0.
$
Finally, we use   the saddle-point approximation (\ref{SPE}) again,  in order to eliminate the cross-product between the terms $(m_i \ddot {\bf x}_i+m_i \gamma_i \dot{\bf x}_i + \nabla_i U)$ and ${\bf F}_i^{bias}$ in the expression for $\Delta S$, yielding one more term  $\propto
| {\bf F}_i^{bias}|^2$. This leads to our final variational condition:
\be\label{VMDcondition}
\frac{\delta}{\delta  X} \int_0^t d\tau \sum_{i=1}^N \frac{1}{\gamma_i m_i}~| {\bf F}_i^{bias}( X;\tau)|^2 \simeq 0.
\ee
This equation states that the optimum reaction trajectory is that for which the time-averaged square modulus of the bias force is least.  Interestingly, a similar condition was recently derived in the context of optimal control theory \cite{Schutte}. 
We emphasize that the functional in Eq.(\ref{VMDcondition}) is not affected by solvent induced fluctuations. 

Let us now extend this discussion to include the case of a  history-dependent biasing force. In particular, we focus on the  ratchet-and-pawl molecular dynamics  (rMD)  algorithm developed in Refs. \cite{Paci1999, Camilloni2011}. The advantage of this formalism is that the bias only sets in whenever the system attempts to backtrack towards the reactant -- defined in terms of some position-dependent reaction coordinate (RC) $z$~--. Conversely, no bias is applied whenever the system spontaneously takes a step towards the product. 

To define the rMD we consider the Langevin equations (\ref{Lang}) with an additional biasing force ${\bf F}^{rMD}_i$  defined as
\be \begin{cases}
      - \frac{k_R}{2}\nabla_i z(X)\cdot ( z(X)- z_m(t) ) & z(X) > z_m(t)\\
       0,  & z(X) \leq z_m(t).
   \label{VR}
\end{cases}
\ee
$z_m(t)$ denotes the smallest value assumed by the RC $z$ up to time~$t$ (we assume that $z$ is minimum in the target), hence obeys the equation of motion
$ \dot z_m =  \dot z \cdot \theta(z_m -z).
 $

 Let us now derive the PI expression for the path probability density  $\mathcal{P}_{rMD}[ X]$. To this end,  we add a small stochastic noise to turn the equation of motion of $z_m$ into an overdamped  Langevin equation.
The PI representation for the path probability density in the extended Langevin system $(X,Y,z_m)$ is readily obtained. Finally, $\mathcal{P}_{rMD}[ X]$  is recovered by taking the small-noise limit and is given by: 
\be\label{PIr1}
\mathcal{P}_{rMD}[ X] &=& \int_{z(X_i)} \mathcal{D} z_m \int \mathcal{D}Y e^{-S_{rMD}[ X,Y, z_m]-\frac{U(X_i,Y_i)}{ k_B T}}\nn\\
&\cdot &~\delta\left[\dot z_m - \dot z[ X] \,\theta(z_m[ X] -z)\right],
\ee 
where $S_{rMD}[X,Y,z_m]$  is obtained from Eq. (\ref{StrVMD}) by setting ${\bf F}^i_{bias}(X,t)={\bf F}^i_{rMD}(X,z_m)$.
From here on, the derivation of the variational principle (\ref{VMDcondition}) is basically identical to the case of an external biasing force reported above (see Supplementary Material, SM).

 \begin{figure}[t!]
\begin{center}
\includegraphics[width= 9 cm]{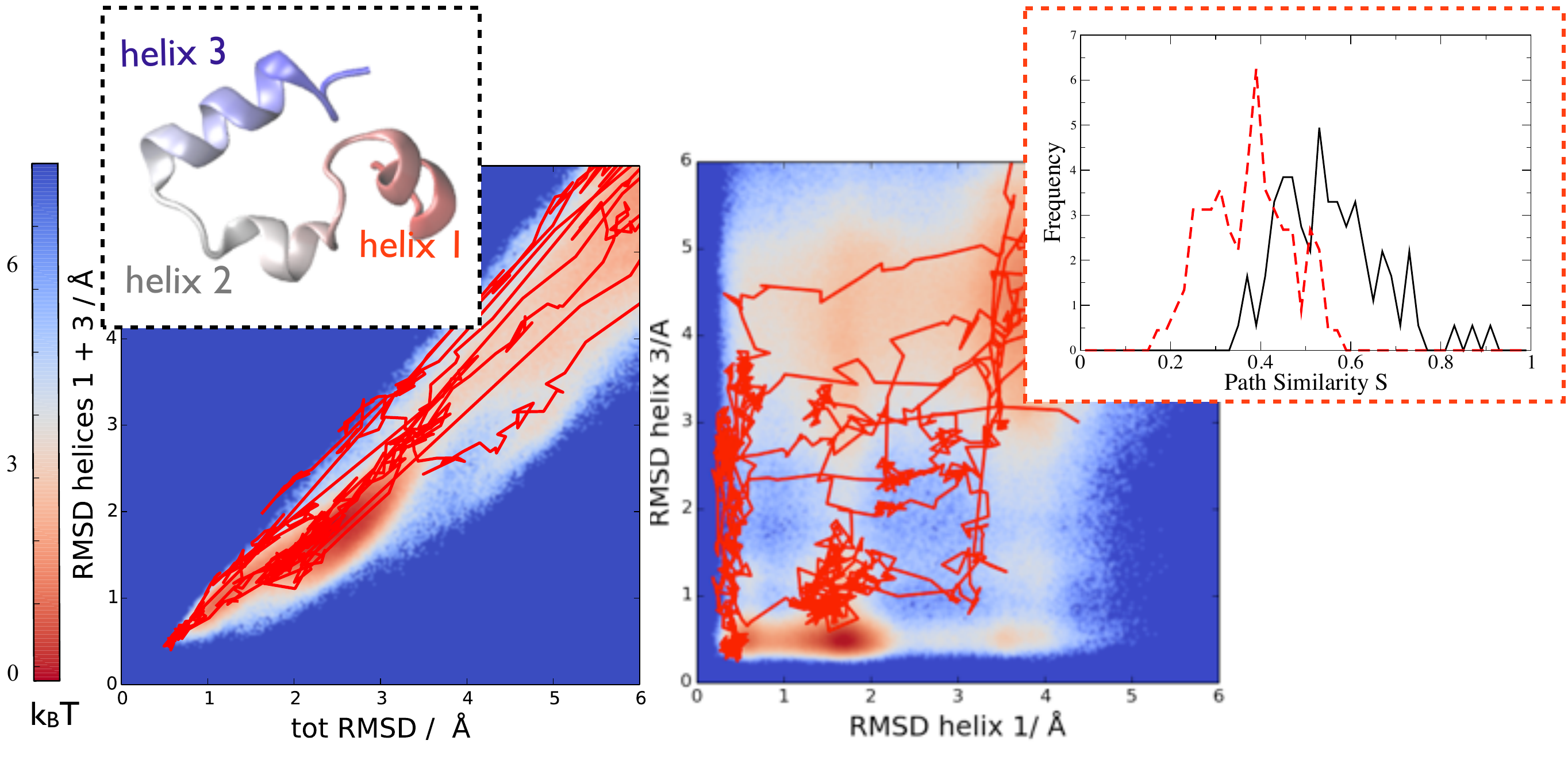}\\
\caption{
Folding trajectories for villin (crystal structure shown in left inset) obtained in the variational approximation, projected on the plain defined by the total RMSD to the native structure and by the RMSD to   the native structure of the residues in helix I and helix III (left panel) and on the plane defined by the RMSD to  the native structure of helix I vs that of helix III (right panel). In the background the free-energy landscape obtained from the Anton MD simulations is shown. In the right inset:  Distribution of similarity between variational and MD folding pathways (dashed line) compared with the intrinsic similarity of MD folding pathways (solid line). }
\label{Fig2}
\end{center}
\end{figure}

Before presenting the results of atomistic protein folding simulations,  it is instructive to illustrate and validate the present variational approximation on a simple toy model, that can be straightforwardly solved on a regular desktop computer. To this end, in the SM we present our study of a transition performed by a point particle diffusing on an asymmetric two-dimensional funnelled energy landscape. The diffusion from the top to the bottom of the funnel is thermally activated, due to the presence of an energy barrier. Plain MD simulations show that the particle reaches the bottom  of the funnel by passing through a ÒgateÓ, i.e. a spatially localised depression on the energy barrier (see Fig. \ref{S1} in the SM). 

We compared the results obtained using different algorithms to generate the trial paths (rMD and standard steered MD) and different values of the biasing force constant $k_R$. In all cases, we chose to bias the dynamics along a rather poor reaction coordinate, which does not take into account the presence of the gate. 

We found that all the trajectories generated  by steered MD very closely follow  the direction selected by the biasing coordinate, hence fail to predict the passage through the gate. Hence, in general, we expect a variational calculation based  on steered MD trial paths to yield rather poor results, unless the reaction coordinate is very accurately known. 

Results obtained by using rMD trial paths are definitely   better (see Fig. \ref{S2} of the SM). In particular, even when choosing a large value for $k_R$, a significant fraction of the trial paths access the bottom of the funnel  through the gate. This is because in rMD  the biasing force is not continuously pushing the system, but only sets in to hinder backtracking. We also note that the variational principle systematically discards unphysical trial rMD trajectories, and correctly predicts the essential qualitative features of the reaction. We conclude the variational calculations based on rMD may yield reasonable results, even when the reaction coordinate is rather poorly known.  

Let us now report our application to the folding transition of two globular proteins: the WW-domain Fip35, (with $\beta$-type native secondary structures, see Fig. 1), and the villin headpiece subdomain (with $\alpha-$type native secondary structures, see Fig. 2). In both cases, we have used the AMBER99SB-ILDN all-atom force field in TIP3P explicit water\cite{Amber}.  Several reversible folding-unfolding MD trajectories for these proteins generated on  \emph{Anton} by using the same force field have been made available by DES Research.

The rMD bias in Eq. (\ref{VR}) was based on the RC introduced in Ref. \cite{Camilloni2011}  (also reported in the SM), defined as the  distance between the instantaneous contact-map and the native state's contact map. The $k_R$ constant was set to $5\times 10^{-3}$~kJ/mol. With this value, the modulus of the total bias force was on average about two orders of magnitude smaller than that of the total physical force.   We tested the robustness of our predictions by repeating the variational calculation with different values of $K_R$ for a given initial condition (see Fig.\ref{S4} in the SM).

For each test protein, we have used the rMD algorithm to produce in total about  one thousand  600~ps-long trial folding trajectories, started from 10 different denatured configurations $X_i^{(1)}, \ldots, X_i^{(10)}$.   The 10 initial conditions were obtained by  1~ns of plain MD at the temperature $T=800~K$, starting from the crystal native state and thermalized by 200~ps at 300~K. 
Folding events were defined as those attaining a final root-mean-square-deviation (RMSD) to the native structure smaller than $2$~\AA.  For each initial condition a single folding trajectory was selected out the ensemble of trial paths by applying condition (\ref{VMDcondition}).

In order to define a convergence criterion for the variational search  we note that the least value of the functional (\ref{VMDcondition}) is non-negative and vanishes for spontaneous transitions. These events have a negligible probability to be observed  in the short simulation time, $t\sim 200$~ps. Typically, we observed that the least value of the functional (\ref{VMDcondition}) decreases on increasing the number of trial trajectories, until it reaches a plateau for more than $\sim50$ trial paths (see Fig. 3). 

It is important to check that, once the plateau region is reached, the predicted folding mechanism does not change when increasing the number of trial trajectories. To this end, we adopted a simplified representation of the folding mechanism realized in a given trajectory: We define a matrix $\hat M$, which describes the order in which the native contacts are formed~\cite{Camilloni2011}.  Namely, let $i,j$ be two indexes running over all native contacts between $C_\alpha$ atoms, and let $t_i(k)$ and $t_j(k)$ be the times  at which they are formed.  The matrix element $M_{i j}(k)$ is defined to be $1(0)$ when  $t_i(k)<t_j(k)$ ($t_i(k)>t_j(k)$) and $1/2$ when $t_i(k)=t_j(k)$.
A quantitative measure of the difference in the folding mechanism followed by two given trajectories  $k$ and $k'$ is provided by their  path similarity $s(k,k')$, defined as
$
s(k,k') = \frac{1}{N_c(N_c-1)}\sum_{i\ne j} \delta(M_{ij}(k)-M_{ij}(k')).
$
Notice that $s(k,k')=1$ if all native contacts are formed in the same order in $k$ and in $k'$, and is $0$ if they are formed in a completely different order.

The  path similarity  can be used to assess the stability of the predicted  folding mechanism in the plateau region. For each given initial condition $X^{(i)}$ we computed the similarity between pairs of variational folding pathways,  obtained using  a different number of trial trajectories.
Namely, we computed the similarity of the variational path obtained with $16$ and $48$, with $48$ and $64$, and with $64$ and $96$ trial trajectories. We found that the mechanism remains stable ($s(k,k') \gtrsim 0.9$) above 48 trial paths, i.e. in the plateau region.   

In Fig.1 we project the folding trajectories for the WW-domain obtained with our variational approach onto the plane defined by the root-mean-square deviation (RMSD) of the two hairpins to the native state and we compare it with the free-energy landscape obtained from a frequency histogram of the long MD trajectories reported in Ref.\cite{DESRes1}.  

Some comments on these results are in order. First, we note that the initial conditions used in the variational calculation are typically  more denatured than the configurations in the equilibrium unfolded state obtained in the Anton simulation. 
In spite of this difference, the variational trajectories reach the native state by traveling along  regions of low free-energy. This fact indicates that the two methods yield the same folding mechanism, i.e. predict that the formation of the secondary structures predominantly occurs in a definite sequence and that in the most likely mechanism the $N$ terminal hairpin folds before the $C$ terminal \cite{DESRes1,Krivov2011}, in agreement with the $\phi$-values analysis of Ref.~\cite{Weikl}. 

To provide a  quantitative measure of the agreement between the variational and the MD paths, we employed again a path similarity analysis.
First, we computed the distribution of $s(k,k')$ within the ensemble of MD folding trajectories (see dashed line in the inset of Fig. 1), to quantify the intrinsic degree of heterogeneity of the folding mechanism. Next, we computed the similarity between all MD and all variational paths, i.e. $s(k, k')$, where $k$ and $k'$ run over MD and variational trajectories, respectively (solid line). The overlap of the two curves  indicates that the average difference between the folding mechanism obtained in the two methods lies within intrinsic statistical fluctuations.
 \begin{figure}[t!]
\begin{center}
\includegraphics[width=6cm]{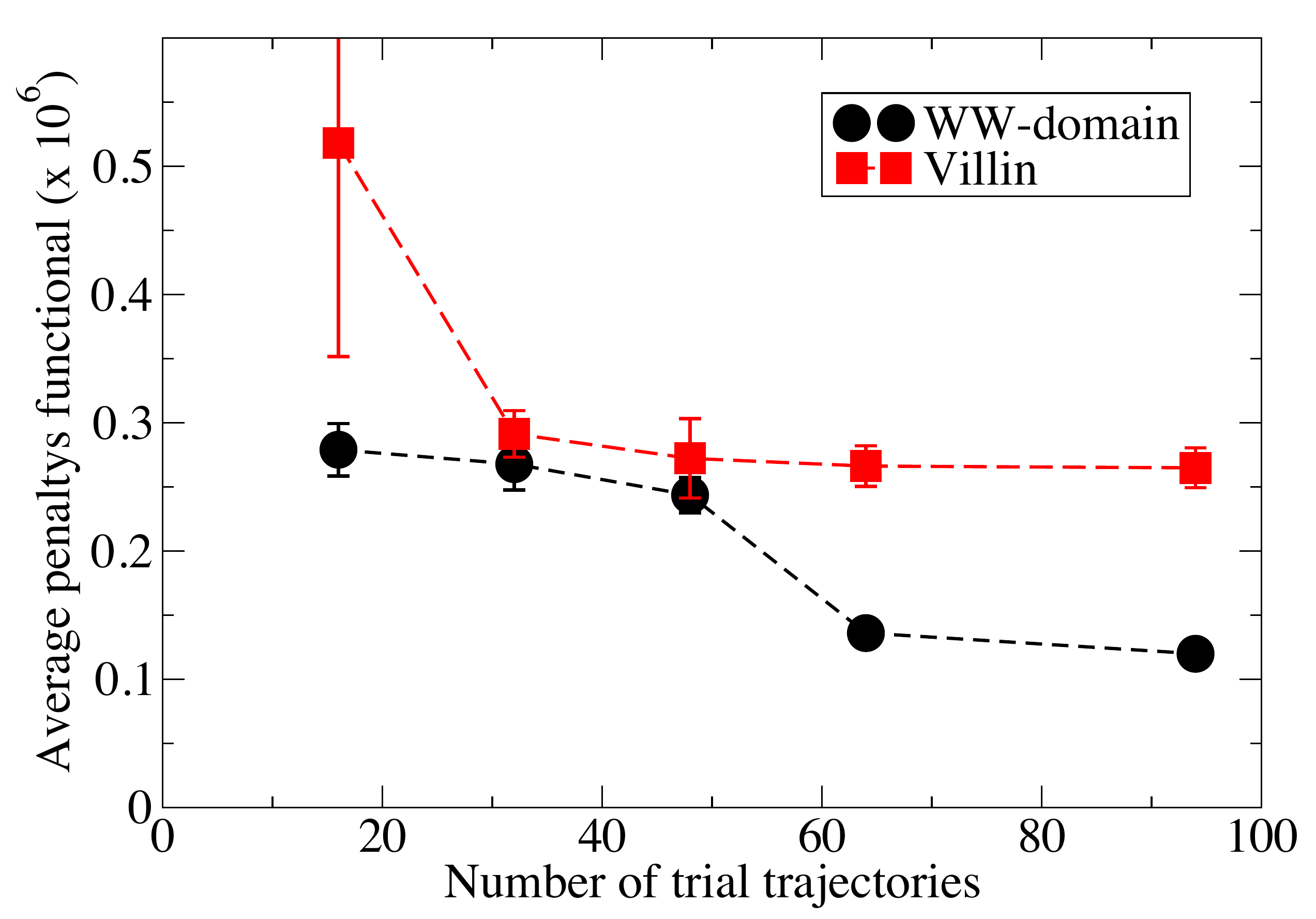}
\caption{The average value of the penalty functional given in Eq. (\ref{VMDcondition}) as a function of the number of trial trajectories.
The average is performed over the different initial conditions.}
\label{Fig3}
\end{center}
\end{figure}

A concern about the variational approach is that the bias may over-promote the rate of formation of local secondary structures, in particular $\alpha-$helixes, relative to that of tertiary structures. In order to test if this is the case, we have studied the folding of villin head-piece subdomain, which contains three $\alpha$-helixes.    
In the left panel of Fig.2 we report our variational folding trajectories projected onto the plane defined by the RMSD to native of the two largest $\alpha-$helixes and the total RMSD to the native structure.  We see that the two  approaches give consistent results and predict that the formation of secondary and tertiary contacts is quite co-operative. Hence, we conclude that the bias force does not enhance the folding rate of $\alpha$-helixes. 

In the right panel of Fig. 2, we project the variational trajectories  onto the plane defined by the RMSD to native of the first and third helix, respectively and we compare it with the corresponding equilibrium free-energy landscape. We note again that the variational trajectories travel along low free-energy regions, correctly predicting that the secondary structures form one after the other. However, we found that the preferential order of helix formation is different in the two calculations. This fact is reflected by a small discrepancy in the path similarity between MD and variational trajectories, of the order of the typical spread of the self-similarity distribution of the MD paths (see the inset in the top-right  corner of Fig. 2).  As a reference, the similarity distribution with random sequences of contacts formation for the folding mechanism predicted by our variational method or by MD is sharply peaked around 0.3 (see Fig.\ref{S5} in the SM).

In conclusion, the variational approach introduced in this work yields the microscopic mechanism for reactions as complex as protein folding, using realistic force fields in all-atom detail. 
%These results were obtained  from just a few hundreds of $\sim ~10^2$~ps-long MD simulations. In contrast, obtaining these results from plain MD required months of computing time on the special-purpose \emph{Anton} facility. 
In view of  its computational efficiency, we foresee applications to many transitions that cannot be simulated by plain MD. The possibility of adopting the explicit solvent all-atom model opens the door to the simulation of conformational changes of other biomolecules, notably nucleic acids.

We thank DES Research for making available their MD simulation data and acknowledge discussions with  H. Orland and S. Piana. All calculations were performed on the Kore cluster at the FBK institute. S. a B. acknowledges support by Istituto Nazionale di Fisica Nucleare through the "Supercalcolo" agreement with Fondazione Bruno Kessler. 

\newpage
{\bf \large SUPPLEMENTARY MATERIAL}
\\
\section{Intermediate steps in the derivation of the Variational principle based on the ratchet-and-pawl molecular dynamics (rMD)}
 Eq. (10) is the basis of the variational approach and was explicitly derived in the case of an \emph{externa}l (i.e. path-independent) biasing force.   In this section we provide some details of the derivation of the same condition in the case of the rMD, where the biasing force depends on the previous history of the system.

In the main text, we have shown that in rMD  the probability for the system to go from $X_i$ to $X_f$ along the path $ X(\tau)$ in time $t$ is given by the path integral (PI)
\be\label{PIr1}
\mathcal{P}_{rMD}[ X] &=& \int_{z(X_i)} \mathcal{D} z_m \int \mathcal{D}Y e^{-S_{rMD}[ X,Y, z_m]-\frac{U(X_i,Y_i)}{2 k_B T}}\nn\\
&\cdot &~\delta\left[\dot z_m - \dot z[ X]\, \theta(z_m -z[ X])\right],
\ee 
The probability to perform the same transition in the \emph{unbiased} Langevin dynamics is given in Eq. (4) of the main text. The same probability can  also be  written in a form involving an additional path integral in $z_m(\tau)$, at the expense of introducing a functional-delta: 
\be\label{PIr2}
\mathcal{P}[ X] &=& \int_{z(X_i)} \mathcal{D} z_m \int \mathcal{D}Y e^{-S_{OM}[ X,Y]-\frac{U(X_i,Y_i)}{2 k_B T}}\nn\\
&\cdot &~\delta\left[\dot z_m- \dot z[ X]\, \theta(z_m -z[ X])\right].
\ee 

The expressions (\ref{PIr1}) and (\ref{PIr2}) are almost identical, except for the fact that the latter contains the exponent of the standard Onsager-Machlup functional $S_{OM}[X,Y]$, rather than the corresponding biased functional $S_{rMD}[X,Y]$. This reflects the fact that in the standard Langevin  dynamics, the history of $z_m$ affects neither the solute nor the solvent dynamics.  

Now, in analogy with  Eq. (6) of the main text, we use the standard reweighing trick to obtain:
\be
\mathcal{P}[ X] = \mathcal{P}_{rMD}[ X] \langle e^{-( S_{OM}[ X,Y]-S_{rMD}[ X, Y, z_m))}\rangle_{z_m, Y}
\ee
where now $\langle\cdot \rangle_{z_m, Y} $ denotes the rMD average over the $Y$ and $z_m$ histories.
Now we note that the term at the exponent does not depend on the solute dynamics,
\be
S_{OM}[ X,Y]-S_{rMD}[ X, Y, z_m] \equiv \Delta S[ X, z_m].
\ee
Thus, the average over $Y$ can be dropped, and one finds:
\be
\mathcal{P}[ X] = \mathcal{P}_{rMD}[ X] \langle e^{-\Delta S[X,z_m]}\rangle_{z_m}
\ee

We now compute the functional derivative of $\mathcal{P}[ X]$ with respect to the trial path $ X$. Following the same saddle-point argument used in the main text in the discussion of the case with an external bias force, we neglect the term proportional to $\delta \mathcal{P}_{rMD}/\delta  X$ and use the saddle-point equations of motion:
\be\label{Eq1}
&&m_i \ddot {\bf x}_i+m_i \gamma_i \dot{\bf x}_i + \nabla_i U-{\bf F}_i^{rMD}(X, z_m)\simeq 0 \\
\label{Eq2}
&&\dot z_m- \dot z[X]\,\theta(z_m -z) =0
\ee
to remove the time-derivative terms from the expression of $\Delta S$. We stress that the second of these equations is exact, while the first is only approximate.
 Finally, denoting with $\hat z_m(t)$ the (unique) solution of Eq(\ref{Eq2}) we arrive to the variational principle:
  \be
  \frac{\delta}{\delta  X} \int_0^t d\tau \sum_{i=1}^N \frac{1}{\gamma_i m_i}~| {\bf F}_i^{rMD}( X;\hat z_m[ X])|^2 \simeq 0.
  \ee

\section{Collective coordinate for rMD simulations of protein folding}

Following Ref. [12] of the main text,  in our rMD simulations we have biased the dynamics according to a reaction coordinate (RC) defining a distance between the instantaneous contact map and the native contact map:
\be
 z(X) \equiv \sum_{|i-j|>35}^{N} [ C_{ij}(X) - C_{ij}(X^{\text{native}}) ]^2.
\ee
In this equation, $C_{ij}(X)$ and $C_{ij}(X^{\text{native}})$ are the instantaneous and native contact maps, respectively. 
Their entries are chosen so as to interpolate smoothly between 0 and 1, depending on the relative distance of the atoms $i$ and $j$:  
\be
C_{ij}(X) = \{ 1-(r_{ij}/ r_0 )^6\}/\{ 1-( r_{ij}/ r_0 )^{10}\},
 \ee
 where r${_0}$=7.5~\AA~ is a fixed reference distance. 
The contribution to the bias force due to a pair of atoms specified by the indexes $i$ and $j$ was set to 0 smoothly any time the distance between these atoms was larger than the cut-off distance $r_c=12$\AA.

\section{Illustrative Application in a Toy Model}
In order to illustrate our variational method and highlight its strengths and limitations, it is instructive to apply it to a toy model that can be straightforwardly simulated on a desktop computer. 

\begin{figure}
\begin{center}
\includegraphics[width= 9 cm]{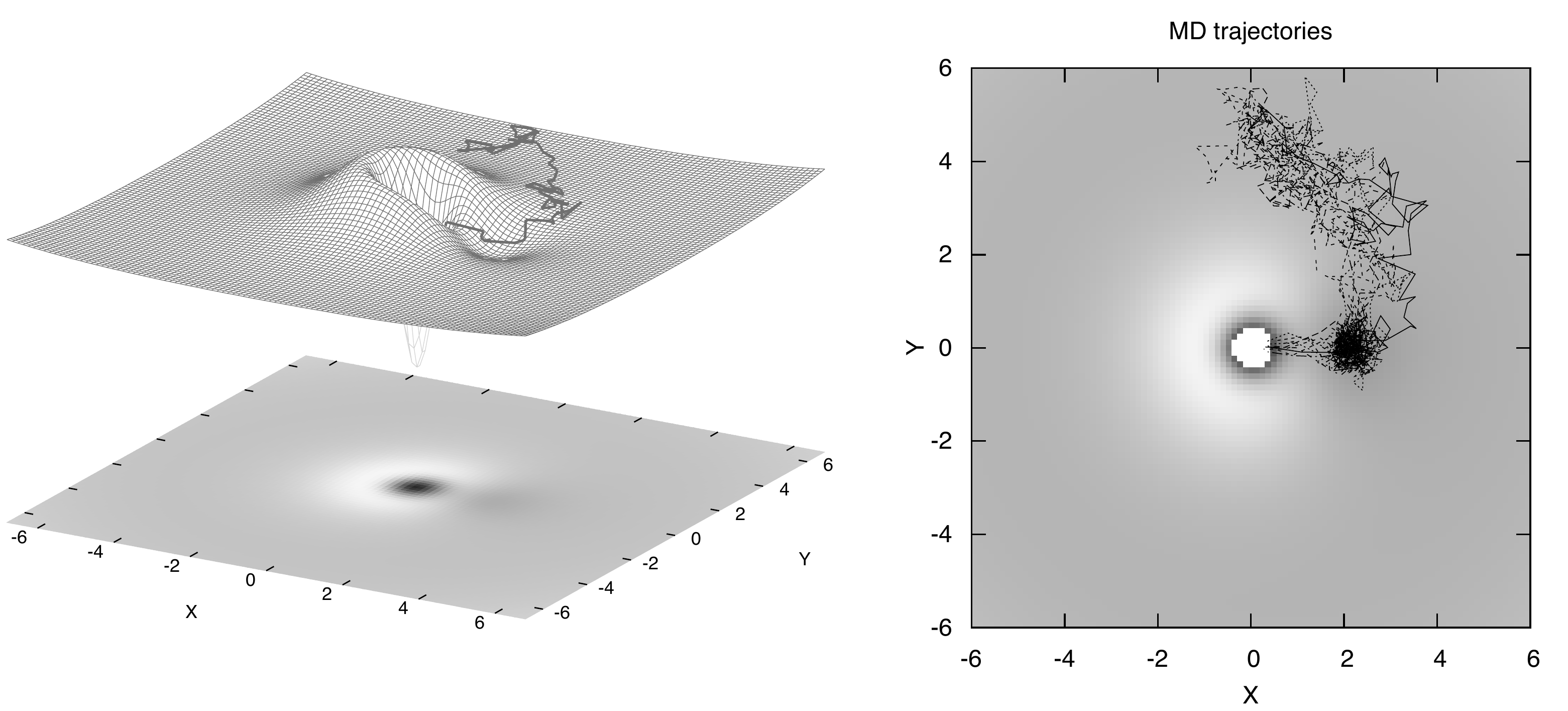}
\caption{Left panel: two dimensional energy surface of the illustrative toy model. The dark line represents a typical reactive trajectory. Right panel: set of reactive trajectories obtained by plain Langevin dynamics. }
\label{S1}
\end{center}
\end{figure}

We consider the diffusion on the two-dimensional energy surface defined by the potential 
\be
U(x,y) &=& w^2 (x^2+y^2)^2  -\frac{A_1 s_1^2}{(x^2+y^2+s_1^2)^2}\nn\\
&&\hspace{-2cm} + \frac{A_2 s_2^2}{(x^2+y^2+s_2^2)^2} - \frac{A_3 s_3^2}{((x-x_m)^2+(y-y_m)^2+s_3^2)^2}\quad
\ee
with $A_1=  30,      A_2=20,       A_3=6$, 
$ s_1=1, s_2=2, s_3=2$,  $w=0.03$, $y_m=0$ and $x_m=1.5$. The corresponding energy landscape is shown in the left panel of Fig.\ref{S1}.

We generated $20,000$ independent  trajectories integrating the  standard underdamped Langevin equation starting  from the same initial condition $(x_i=0, y_i=5)$ located in the outer flat region. We used $\gamma=1, dt=0.02$ and $k_BT=0.2$, and selected only the paths which reached the product state $(x=0,y=0)$,  at the bottom of the funnel.

At this temperature, crossing the ring barrier surrounding the funnel is a thermally activated process. 
As a result, after 90,000 integration steps only a few trajectories reached the product state, by accessing the funnel through the gate located at $(x \sim 1.5, y\sim0)$ (see right panel of Fig.\ref{S1}). 
 
Let us now compare these results with those obtained using our variational approach. As a first step,  we integrated  2000 trial trajectories of rMD dynamics consisting of 30,000 time-steps,  using as biasing coordinate the Euclidean distance from the product,  $z = \sqrt{x^2+y^2}$. We emphasise that this choice of reaction coordinate is certainly not optimal, since it does not take into account  the existence of the gate. 

Next, we selected the optimum trajectory among all the trial trajectories, by applying the variational condition given in Eq. (8).  
We repeated this procedure for a wide range of biasing strength constants $K_R$, covering over more than two orders of magnitude. 
We emphasize that also in this model  rMD allows to generate a large ensemble of trial reactive trajectories at a fraction of the computational cost required by plain Langevin dynamics.  

\begin{figure}
\begin{center}
\includegraphics[width= 9 cm]{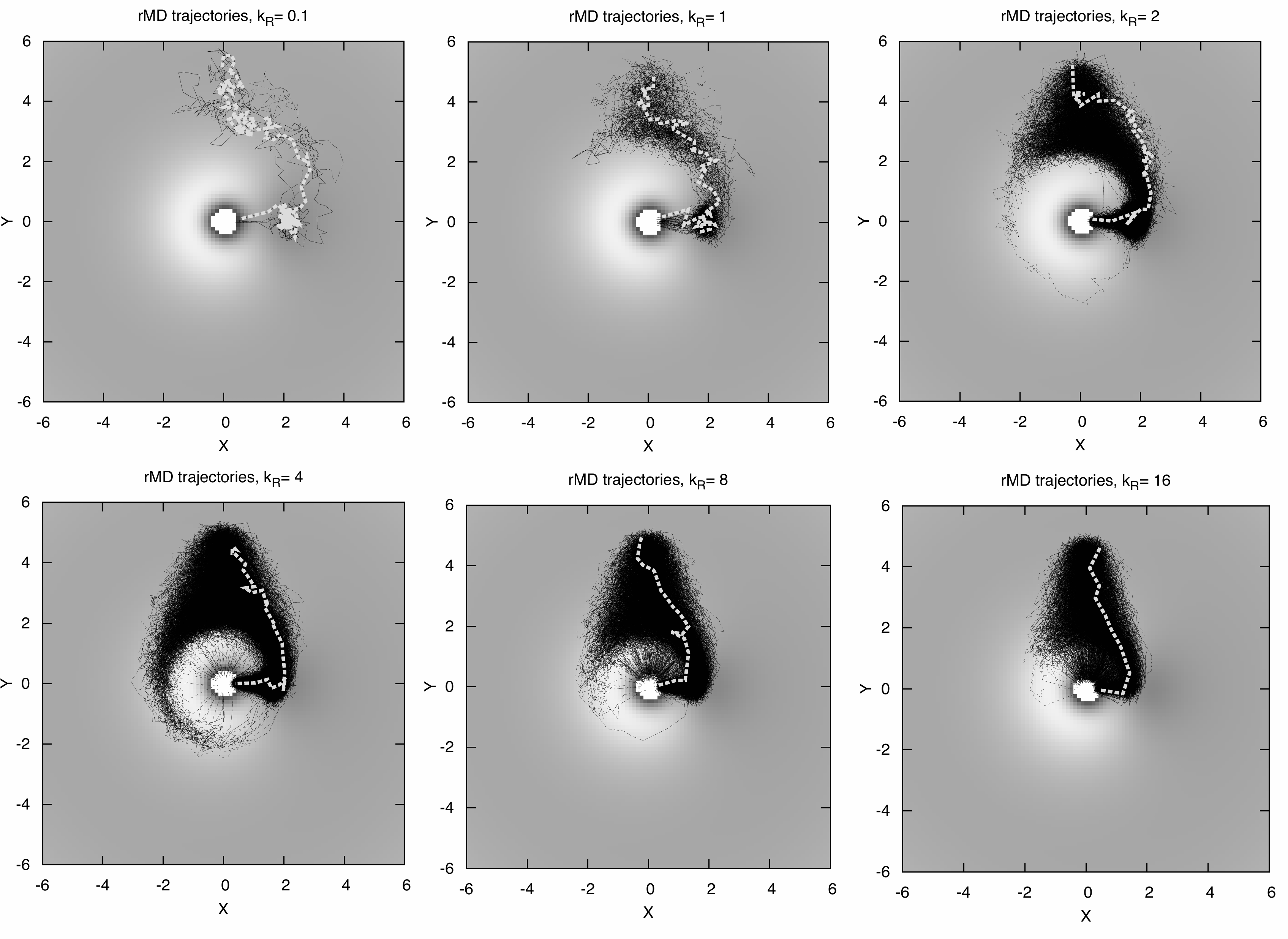}
\caption{Reactive pathways obtained by the variational approach using different values of the biasing constant $K_R$. The light line is the optimum, selected according to the variational condition (8). }
\label{S2}
\end{center}
\end{figure}

The results are shown in Fig. \ref{S2}. First we note that, at low values of the biasing force ($K_R< 4$),  the fraction of trial trajectories reaching the product state within the simulation time varies significantly with $K_R$. For larger values of $K_R$, saturation is reached and almost all trial trajectories attain the final state.

As $K_R$ is raised, we observe an increasing  fraction of trial reactive  trajectories crossing the ring barrier, in contrast with what is seen in plain Langevin dynamics. This is an artifact due to our bad choice of the biasing reaction coordinate.
However, once the variational condition is applied, such unrealistic paths are discarded.  Indeed,  the results obtained at all values of $K_R$ correctly predict the essential feature of the reaction, i.e. the crossing the gate. On the other hand,  the optimum paths corresponding to the largest values of biasing strength are significantly shorter and  tend to travel along a line closer to the vertical $\hat y$ axis.

\begin{figure}
\begin{center}
\includegraphics[width= 7 cm]{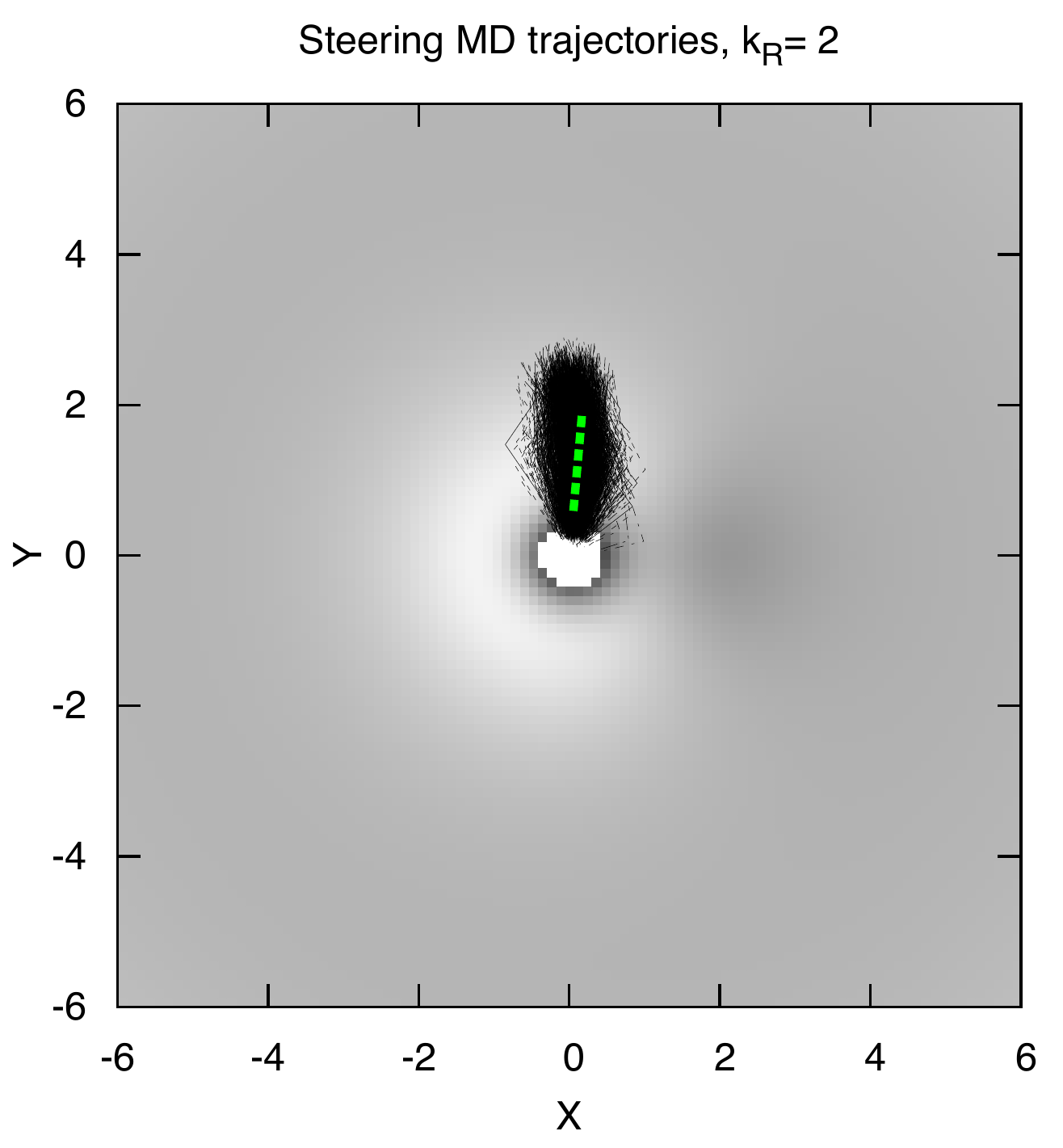}
\caption{Variational calculation of the reaction pathways obtained using a steered MD algorithm (instead of the rMD algorithm) to generate the trial paths. }
\label{S3}
\end{center}
\end{figure}

Finally, it is interesting to compare the results obtained using rMD to generate the trial paths to those obtained using a standard steered MD, where a constant harmonic force with strength  constant $K_R$ is introduced to guide the path towards the bottom of the funnel. The results shown in Fig. \ref{S3}
clearly show that the performance of the variational method is significantly worse for calculations based on steered MD than on rMD. The better performance of rMD as compared to steered MD is probably due to the fact that former scheme does not push the system towards the product state, but it only hinders backtracking. This feature also reduces the effect of a suboptimal choice of the reaction coordinate.

\section{Dependence on the folding pathways on the strength of the biasing force}
\begin{figure}
\begin{center}
\includegraphics[width= 7 cm]{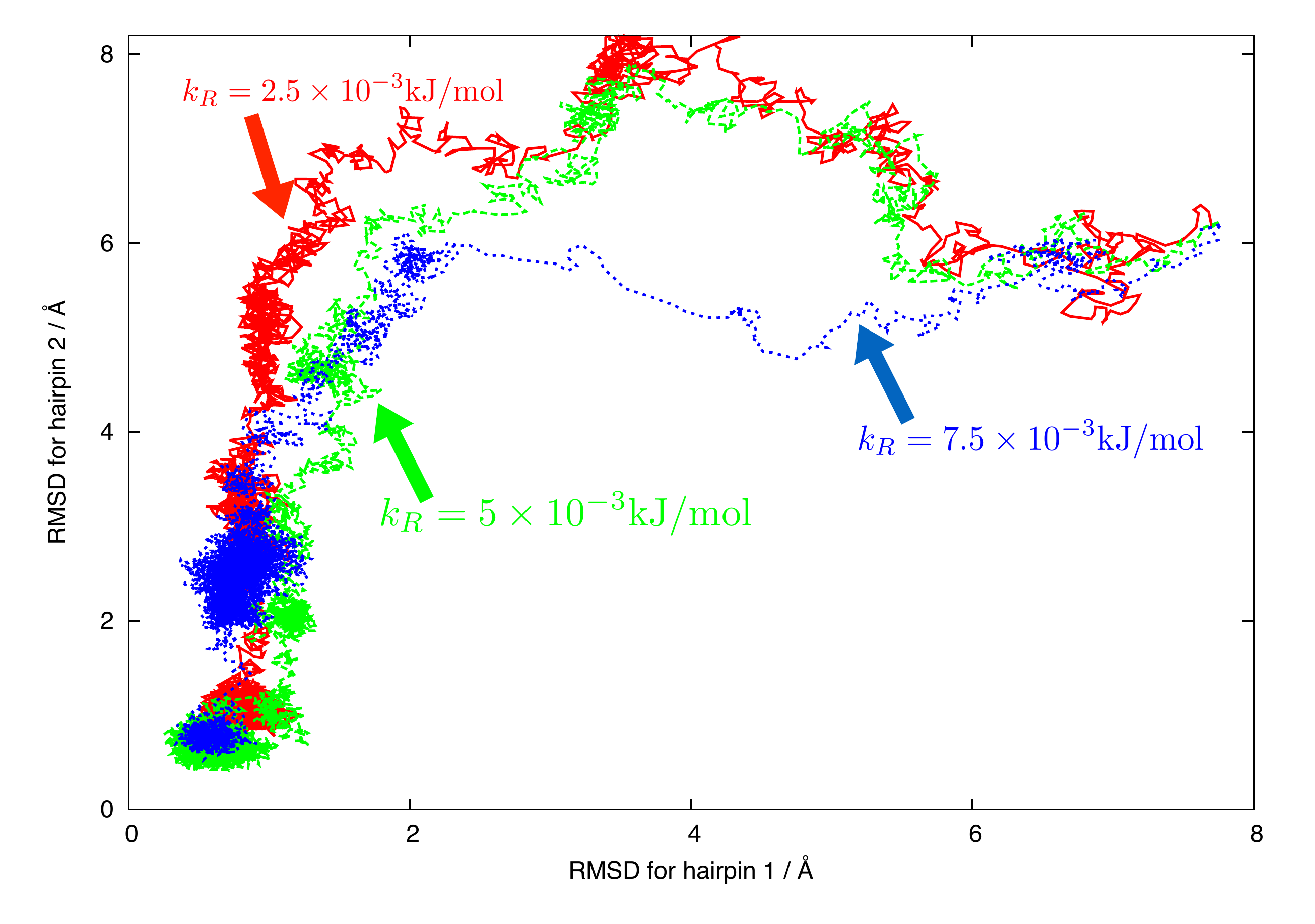}
\caption{Comparison between the reaction pathways for Fip35 obtained starting from the same initial conditions using three different values for  the strength of the biasing constant $K_R= 2.5, 5.0, 7.5 \times 10^{-3}$~kJ/mol. }
\label{S4}
\end{center}
\end{figure}

To assess the sensitivity of the atomistic variational results on the choice of the biasing constant $k_R$, in Fig.\ref{S4} we compare the results obtained for the folding of Fip35 from a given initial condition, using  $K_R= 2.5, 5, 7.5 \times 10^{-3}$~kJ/mol. We find that the resulting folding mechanism is the same. 

\section{Path similarity against random native contact formation}
\begin{figure}
\begin{center}
\includegraphics[width= 7 cm]{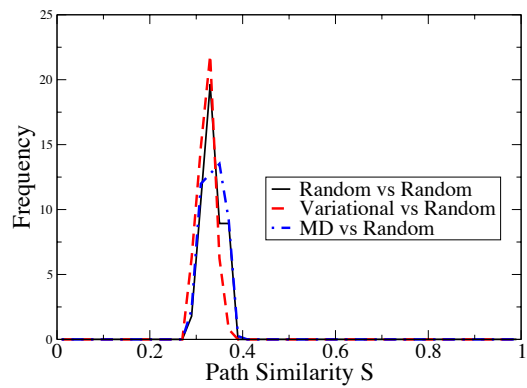}
\caption{Distribution of similarity between the order of native contact formation in  villin obtained in different theoretical approaches and in an ensemble of random sequences  of  native contact formation. The solid line denotes the distribution of similarity  between two random sequences of native contact formation, the dashed (dot-dashed) line the similarity between variational (plain MD) trajectories and random sequences of native contact formation. }
\label{S5}
\end{center}
\end{figure}

In Fig.\ref{S5} we show the distribution of  similarity between the orders of contact formation found in the folding pathways calculated for villin using the variational approach or plain MD and a random order of contact formation.  As a reference, the solid line denotes the distribution of similarity between pairs of random sequences of native contact formations. 
We see that these distributions are indistinguishable and are sharply peaked around 0.3. 
\newpage

\end{document}